\documentclass[aps,preprint]{revtex4}%
\usepackage{amsfonts}
\usepackage{amsmath}
\usepackage{amssymb}
\usepackage{graphicx}%
\begin{document}
\title[Random matrix theory within superstatistics]{Random matrix theory within superstatistics}
\author{A. Y. Abul-Magd}
\keywords{Random matrix theory}
\pacs{05.40.-a, 05.45.Mt, 03.65.-w, 02.30.Mv}

\begin{abstract}
We propose a generalization of the random matrix theory following the basic
prescription of the recently suggested concept of superstatistics. Spectral
characteristics of systems with mixed regular-chaotic dynamics are expressed
as weighted averages of the corresponding quantities in the standard theory
assuming that the mean level spacing itself is a stochastic variable. We
illustrate the method by calculating the level density, the
nearest-neighbor-spacing distributions and the two-level correlation functions
for system in transition from order to chaos. The calculated spacing
distribution fits the resonance statistics of random binary networks obtained
in a recent numerical experiment.

\end{abstract}
\date[Date text]{\today}
\startpage{1}
\endpage{2}
\maketitle

\section{Introduction}

Random matrix theory (RMT) provides a suitable framework to describe quantal
systems whose classical counterpart has a chaotic dynamics \cite{mehta,haake}.
It models a chaotic system by an ensemble of random Hamiltonian matrices $H$
that belong to one of the three universal classes, namely the Gaussian
orthogonal, unitary and symplectic ensembles (GOE, GUE and GSE). The theory is
based on two main assumptions: (i) the matrix elements are independent
identically-distributed random variables, and (ii) their distribution is
invariant under unitary transformations. These lead to a Gaussian probability
density distribution for the matrix elements, $P\left(  H\right)
\varpropto\exp\left[  -\eta\text{Tr}\left(  H^{\dagger}H\right)  \right]  $.
With these assumptions, RMT presents a satisfactory description for numerous
chaotic systems. On the other hand, there are elaborate theoretical arguments
by Berry and Tabor \cite{tabor}, which are supported by several numerical
calculations, that the nearest-neighbor-spacing (NNS) distribution of
classically integrable systems should have a Poisson distribution $\exp(-s)$,
although exceptions exist.

For most systems, however, the phase space is partitioned into regular and
chaotic domains. These systems are known as mixed systems. Attempts to
generalize RMT to describe such mixed systems are numerous; for a review
please see \cite{guhr}. Most of these attempts are based on constructing
ensembles of random matrices whose elements are independent but not
identically distributed. Thus, the resulting expressions are not invariant
under base transformation. The first work in this direction is due to
Rosenzweig and Porter \cite{rosen}. They model the Hamiltonian of the mixed
system by a superposition of a diagonal matrix of random elements having the
same variance and a matrix drawn from a GOE. Therefore, the variances of the
diagonal elements of the total Hamiltonian are different from those of the
off-diagonal ones, unlike the GOE Hamiltonian in which the variances of
diagonal elements are twice those of the off-diagonal ones. Hussein and Pato
\cite{hussein} used the maximum entropy principle to construct "deformed"
random-matrix ensembles by imposing different constraints for the diagonal and
off-diagonal elements. This approach has been successfully applied to the case
of metal-insulator transition\cite{hussein1}. A recent review of the deformed
ensemble is given in \cite{hussein2}. Ensembles of band random matrices, whose
entries are equal to zero outside a band of limited width along the principal
diagonal, have often been used to model mixed systems
\cite{casati,fyodorov,haake}. However, so far in the literature, there is no
rigorous statistical description for the transition from integrability to
chaos. The field remains open for new proposals.

The past decade has witnessed a considerable interest devoted to the possible
generalization of statistical mechanics. Much work in this direction followed
Tsallis seminal paper \cite{Ts1}. Tsallis introduced a non-extensive entropy,
which depends on a positive parameter$\ q$ known as the entropic index. The
standard Shannon entropy is recovered for $q$ = 1. Applications of the Tsallis
formalism covered a wide class of phenomena; for a review please see, e.g.
\cite{Ts2}. Recently, the formalism has been applied to include systems with
mixed regular-chaotic dynamics in the framework of RMT
\cite{evans,toscano,bertuola,nobre,abul1,abul2}. This is done by extremizing
Tsallis' non-extensive entropy, rather than Shannon's, but again subject to
the same constraints of normalization and existence of the expectation value
of Tr$\left(  H^{\dagger}H\right)  $. The latter constraint preserves base
invariance. The first attempt in this direction is probably due to Evans and
Michael \cite{evans}. Toscano et al. \cite{toscano} constructed non-Gaussian
ensemble by minimizing Tsallis' entropy and obtained expressions for the level
densities and spacing distributions for mixed systems belonging to the
orthogonal-symmetry universality class. Bertuola et al. \cite{bertuola}
expressed the spectral fluctuation in the subextensive regime in terms of the
gap function, which measures the probability of an eigenvalue-free segment in
the spectrum. A slightly different application of non-extensive statistical
mechanics to RMT is due to Nobre et al. \cite{nobre}. The
nearest-neighbor-spacing (NNS) distributions obtained in this approach decays
as a power-law for large spacings. Such anomalous distributions can hardly be
used to interpolate between nearly-regular systems which have almost
exponential NNS distributions and nearly-chaotic ones whose distributions
behave at large spacing as Gaussians. Moreover, the constraints of
normalization and existence of an expectation value for Tr$\left(  H^{\dagger
}H\right)  $ set up an upper limit for the entropic index $q$ beyond which the
involved integrals diverge. This restricts the validity of the non-extensive
RMT to a limited range\ near the chaotic phase \cite{abul1,abul2}.

Another extension of statistical mechanics is provided by the formalism of
superstatistics (statistics of a statistics), recently proposed by Beck and
Cohen \cite{BC}. Superstatistics arises as weighted averages of ordinary
statistics (the Boltzmann factor) due to fluctuations of one or more intensive
parameter (e.g. the inverse temperature). It includes Tsallis' non-extensive
statistics, for $q\geq1$, as a special case in which the inverse temperature
has a $\chi^{2}$-distributions. With other distributions of the intensive
parameters, one comes to other more general superstatistics. Generalized
entropies, which are analogous to the Tsallis entropy, can be defined for
these general superstatistics \cite{abe,souza,souzaTs}. This formalism has
been elaborated and applied successfully to a wide variety of physical
problems, e.g., in
\cite{cohen,beck,beckL,salasnich,sattin,reynolds,ivanova,beckT}.

In a previous paper \cite{supst}, the concept of superstatistics was applied
to model a mixed system within the framework of RMT. The joint matrix element
distribution was represented as an average over $\exp\left[  -\eta
\text{Tr}\left(  H^{\dagger}H\right)  \right]  $ with respect to the parameter
$\eta$. An expression for the eigenvalue distributions was deduced. Explicit
analytical results were obtained for the special case of two-dimensional
random matrix ensembles. Different choices of parameter distribution, which
had been studied in Beck and Cohen's paper \cite{BC} were considered. These
distributions essentially led to equivalent results for the level density and
NNS distributions. The present paper is essentially an extension of the
superstatistical approach of Ref. \cite{supst} to random-matrix ensembles of
arbitrary dimension. The distribution of local mean level densities is
estimated by applying the principle of maximum entropy, as done by Sattin
\cite{sattin}. In Section 2 we briefly review the superstatistics concept and
introduce the necessary generalization required to express the characteristics
of the spectrum of a mixed system into an ensemble of chaotic spectra with
different local mean level density. The evolution of the eigenvalue
distribution during the stochastic transition induced by increasing the
local-density fluctuations is considered in Section 3. The corresponding NNS
distributions are obtained in Section 4 for systems in which the time-reversal
symmetry is conserved or violated. Section 5 considers the two-level
correlation functions. The conclusion of this work is formulated in Section 6.

\section{FORMALISM}

\subsection{Superstatistics and RMT}

To start with, we briefly review the superstatistics concept as introduced by
Beck and Cohen \cite{BC}. Consider a non-equilibrium system with
spatiotemporal fluctuations of the inverse temperature $\beta$. Locally, i.e.
in spatial regions (cells) where $\beta$ is approximately constant, the system
may be described by a canonical ensemble in which the distribution function is
given by the Boltzmann factor $e^{-\beta E}$, where $E$ is an effective energy
in each cell. In the long-term run, the system is described by an average over
the fluctuating $\beta$. The system is thus characterized by a convolution of
two statistics, and hence the name \textquotedblright
superstatistics\textquotedblright. One statistics is given by the Boltzmann
factor and the other one by the probability distribution $f(\beta)$ of $\beta$
in the various cells. One obtains Tsallis' statistics when $\beta$ has a
$\chi^{2}$ distribution, but this is not the only possible choice. Beck and
Cohen give several possible examples of functions which are possible
candidates for $f(\beta)$. Sattin \cite{sattin} suggested that, lacking any
further information, the most probable realization of $f(\beta)$ will be the
one that maximizes the Shannon entropy. Namely this version of superstatistics
formalism will now be applied to RMT.

Gaussian random-matrix ensembles have several common features with the
canonical ensembles. In RMT, the square of a matrix element plays the role of
energy of a molecule in a gas. When the matrix elements are statistically
identical, one expects them to become distributed as the Boltzmann's. One
obtains a Gaussian probability density distribution of the matrix elements
\begin{equation}
P\left(  H\right)  \varpropto\exp\left[  -\eta\text{Tr}\left(  H^{\dagger
}H\right)  \right]
\end{equation}
by extremizing the Shannon entropy \cite{mehta,balian} subjected to the
constraints of normalization and existence of the expectation value of
Tr$\left(  H^{\dagger}H\right)  $. The quantity\ Tr$\left(  H^{\dagger
}H\right)  $\ plays the role of the effective energy of the system, while the
role of the inverse temperature $\beta$ is played by $\eta$, being twice the
inverse of the matrix-element variance.

Our main assumption is that Beck and Cohen's superstatistics provides a
suitable description for systems with mixed regular-chaotic dynamics. We
consider the spectrum of a mixed system as made up of many smaller cells that
are temporarily in a chaotic phase. Each cell is large enough to obey the
statistical requirements of RMT but has a different distribution parameter
$\eta$ associated with it, according to a probability density $\widetilde
{f}(\eta)$. Consequently, the superstatistical random-matrix ensemble that
describes the mixed system is a mixture of Gaussian ensembles. Its
matrix-element joint probability density distributions obtained by integrating
distributions of the form in Eq. (1) over all positive values of $\eta$\ with
a statistical weight $\widetilde{f}(\eta)$,
\begin{equation}
P(H)=\int_{0}^{\infty}\widetilde{f}(\eta)\frac{\exp\left[  -\eta
\text{Tr}\left(  H^{\dagger}H\right)  \right]  }{Z(\eta)}d\eta,
\end{equation}
where $Z(\eta)=\int\exp\left[  -\eta\text{Tr}\left(  H^{\dagger}H\right)
\right]  d\eta$. Here we use the \textquotedblright B-type
superstatistics\textquotedblright\ \cite{BC}. The distribution in Eq. (2) is
isotropic in the matrix-element space. Relations analogous to Eq. (1) can also
be written for the joint distribution of eigenvalues as well as any other
statistic that is obtained from it by integration over some of the
eigenvalues, such as the nearest-neighbor-spacing distribution and the level
number variance. The distribution $\widetilde{f}(\eta)$ has to be
normalizable, to have at least a finite first moment
\begin{equation}
\left\langle \eta\right\rangle =\int_{0}^{\infty}\widetilde{f}(\eta)\eta
d\eta,
\end{equation}
and to be reduces a delta function as the system becomes fully chaotic.

The random-matrix distribution\ in Eq. (2) is invariant under base
transformation because it depends on the Hamiltonian matrix elements through
the base-invariant quantity \ Tr$\left(  H^{\dagger}H\right)  $. Factorization
into products of individual element distributions is lost here, unlike in the
distribution functions of the standard RMT and most of its generalizations for
mixed systems. The matrix elements are no more statistically independent. This
handicaps one in carrying numerical calculations by the random-number
generation of ensembles and forces one to resort to artificial methods as done
in \cite{toscano}. Base invariance makes the proposed random-matrix formalism
unsuitable for description of nearly integrable systems. These systems are
often described by an ensemble of diagonal matrices in a presumably fixed
basis. For this reason we expect the present superstatistical approach to
describe only the final stages of the stochastic transition. The base
invariant theory in the proposed form does not address the important problem
of symmetry breaking in a chaotic system, where the initial state is modelled
by a block diagonal matrix with $m$ blocks, each of which is a GOE
\cite{guhr}. This problem is well described using deformed random-matrix
ensembles as in \cite{hussein} or phenomenologically by considering the
corresponding spectra as superpositions of independent sub-spectra, each
represented by a GOE \cite{aas}.

The physics behind the proposed superstatistical generalization of RMT is the
following. The eigenstates of a chaotic system are extended and cover the
whole domain of classically permitted motion randomly, but uniformly. They
overlap substantially, as manifested by level repulsion. There are no
preferred eigenstate; the states are statistically equivalent. As a result,
the matrix elements of the Hamiltonian \ in any basis are independently but
identically distributed, which leads to the Wigner-Dyson statistics. Coming
out of the chaotic phase, the extended eigenstates become less and less
homogeneous in space. Different eigenstates become localized in different
places and the matrix elements that couple different pairs are no more
statistically equal. The matrix elements will no more have the same variance;
one has to allow each of them to have its own variance. But this will
dramatically increase the number of parameters of the theory. The proposed
superstatistical approach solves this problem by treating all of the matrix
elements as having a common variance, not fixed but fluctuating.

\subsection{Eigenvalue distribution}

The matrix-element distribution is not directly useful in obtaining numerical
results concerning energy-level statistics such as the nearest-neighbor
spacing distribution, the two-point correlation function, the spectral
rigidity, and the level-number variance. These quantities are presumably
obtainable from the eigenvalue distribution. From (1), it is a simple
matter\ to set up the eigenvalue distribution of a Gaussian ensemble. With
$H=U^{-1}EU$, where $U$\ is the global unitary group, we introduce the
elements of the diagonal matrix of eigenvalues $E=$ diag$(E_{1},\cdots,E_{N})$
of the eigenvalues and the independent elements of $U$ as new variables. Then
the volume element (4) has the form
\begin{equation}
dH=\left\vert \Delta_{N}\left(  E\right)  \right\vert ^{\beta}dEd\mu(U),
\end{equation}
where $\Delta_{N}\left(  E\right)  =\prod_{n>m}(E_{n}-E_{m})$ is the
Vandermonde determinant and $d\mu(U)$ the invariant Haar measure of the
unitary group \cite{mehta,guhr}. Here $\beta=1,2$ and 4 for GOE, GUE and GSE,
respectively. The probability density $P_{\beta}(H)$ is invariant under
arbitrary rotations in the matrix space.\ Integrating over $U$ yields the
joint probability density of eigenvalues in the form
\begin{equation}
P_{\beta}(E_{1},\cdots,E_{N})=\int_{0}^{\infty}f(\eta)P_{\beta}^{(G)}%
(\eta,E_{1},\cdots,E_{N})d\eta,
\end{equation}
where $P_{\beta}^{(G)}(\eta,E_{1},\cdots,E_{N})$ is the eigenvalue
distribution of the corresponding Gaussian ensemble, which is given by
\begin{equation}
P_{\beta}^{(G)}(\eta,E_{1},\cdots,E_{N})=C_{\beta}\left\vert \Delta_{N}\left(
E\right)  \right\vert ^{\beta}\exp\left[  -\eta\sum_{i=1}^{N}E_{i}^{2}\right]
,
\end{equation}
where $C_{\beta}$ is a normalization constant. Similar relations can be
obtained for any statistic $\sigma_{\beta}(E_{1},\cdots,E_{k}),$ with
$k<N,$\ that can be obtained from $P_{\beta}(E_{1},\cdots,E_{N})$\ by
integration over the eigenvalues $E_{k+1},\cdots,E_{N}$.

In practice, one has a spectrum consisting of a series of levels $\left\{
E_{i}\right\}  ,$ and is interested in their fluctuation properties. In order
to bypass the effect of the level density variation, one introduces the so
called \textquotedblright unfolded spectrum\textquotedblright\ $\left\{
\varepsilon_{i}\right\}  $, where $\varepsilon_{i}=E_{i}/D$ and $D$ is the
local mean level spacing. Thus, the mean level density of the unfolded
spectrum is unity. On the other hand, the energy scale for a Gaussian
random-matrix ensemble is defined by the parameter $\eta.$ The mean level
spacing may be expressed as%
\begin{equation}
D=\frac{c}{\sqrt{\eta}},
\end{equation}
where $c$ is a constant depending on the size of the ensemble. Therefore,
although the parameter $\eta$ is the basic parameter of RMT, it is more
convenient for practical purposed to consider the local mean spacing $D$
itself instead of $\eta$ as the fluctuating variable for which superstatistics
has to be established.

The new framework of RMT provided by superstatistics should now be clear. The
local mean spacing $D$ is no longer a fixed parameter but it is a stochastic
variable with probability distribution $f(D)$. Instead, the the observed mean
level spacing is just its expectation value. The fluctuation of the local mean
spacing is due to the correlation of the matrix elements which disappears for
chaotic systems. In the absence of these fluctuations, $f(D)=\delta(D-1)$ and
we obtain the standard RMT. Within the superstatistics framework, we can
express\ any statistic $\sigma(E)$ of a mixed system that can in principle
be\ obtained from the joint eigenvalue distribution by integration over some
of the eigenvalues, in terms of the corresponding statistic $\sigma
^{(G)}(E,D)$ for a Gaussian random ensemble. The superstatistical
generalization is given by
\begin{equation}
\sigma(E)=\int_{0}^{\infty}f(D)\sigma^{(G)}(E,D)dD.
\end{equation}
The remaining task of superstatistics is the computation of the distribution
$f(D)$.

\subsection{Evaluation of the local-mean-spacing distribution}

Following Sattin \cite{sattin}, we use the principle of maximum entropy to
evaluate the distribution $f(D)$. Lacking a detailed information about the
mechanism causing the deviation from the prediction of RMT, the most probable
realization of $f(D)$ will be the one that extremizes the Shannon entropy
\begin{equation}
S=-\int_{0}^{\infty}f(D)\ln f(D)dD
\end{equation}
with the following constraints:

\textbf{Constraint 1}. The major parameter of RMT is $\eta$ defined in Eq.
(1). Superstatistics was introduced in Eq. (2) by allowing $\eta$ to fluctuate
around a fixed mean value $\left\langle \eta\right\rangle $. This implies, in
the light of Eq. (7), the existence of the mean inverse square of $D$,
\begin{equation}
\left\langle D^{-2}\right\rangle =\int_{0}^{\infty}f(D)\frac{1}{D^{2}}dD.
\end{equation}

\textbf{Constraint 2}. The fluctuation properties are usually defined for
unfolded spectra, which have a unit mean level spacing. We thus require
\begin{equation}
\int_{0}^{\infty}f(D)DdD=1.
\end{equation}

Therefore, the most probable $f(D)$ extremizes the functional
\begin{equation}
F=-\int_{0}^{\infty}f(D)\ln f(D)dD-\lambda_{1}\int_{0}^{\infty}f(D)DdD-\lambda
_{2}\int_{0}^{\infty}f(D)\frac{1}{D^{2}}dD
\end{equation}
where $\lambda_{1}$ and $\lambda_{2}$ are Lagrange multipliers. As a result,
we obtain
\begin{equation}
f(D)=C\exp\left[  -\alpha\left(  \frac{2D}{D_{0}}+\frac{D_{0}^{2}}{D^{2}%
}\right)  \right]
\end{equation}
where $\alpha$ and $D_{0}$ are parameters, which can be expressed in terms of
the Lagrange multipliers $\lambda_{1}$ and $\lambda_{2}$, and $C$ is a
normalization constant. We determine $D_{0}$ and $C$ by using Eqs. (10) and
(11) as
\begin{equation}
D_{0}=\alpha\frac{G_{03}^{30}\left(  \left.  \alpha^{3}\right\vert 0,\frac
{1}{2},1\right)  }{G_{03}^{30}\left(  \left.  \alpha^{3}\right\vert
0,1,\frac{3}{2}\right)  },
\end{equation}
and
\begin{equation}
C=\frac{2\alpha\sqrt{\pi}}{D_{0}G_{03}^{30}\left(  \left.  \alpha
^{3}\right\vert 0,\frac{1}{2},1\right)  }.
\end{equation}
Here $G_{03}^{30}\left(  \left.  x\right\vert b_{1},b_{2},b_{2}\right)  $ is a
Meijer's G-function defined in the Appendix.

\section{LEVEL DENSITY}

The density of states can be obtained from the joint eigenvalue distribution
directly by integration%
\begin{equation}
\rho(E)=N%
{\displaystyle\idotsint}
P_{\beta}(E,E_{2},\cdots,E_{N})dE_{2}\cdots dE_{N}.
\end{equation}
For a Gaussian ensemble, simple arguments \cite{mehta,porter} lead to Wigner's
semi-circle law%
\begin{equation}
\rho_{\text{GE}}(E,D)=\left\{
\begin{array}
[c]{c}%
\frac{2N}{\pi R_{0}^{2}}\sqrt{R_{0}^{2}-E^{2}},\text{ for }\left\vert
E\right\vert \leq R_{0}\\
0,\text{ \ \ \ \ \ \ \ \ \ \ \ \ \ \ \ \ \ \ \ for }\left\vert E\right\vert
>R_{0}%
\end{array}
\right.  ,
\end{equation}
where $D$ is the mean level spacing, while the prefactor is chosen so that
$\rho_{\text{GE}}(E)$\ satisfies the normalization condition
\begin{equation}%
{\displaystyle\int_{-\infty}^{\infty}}
\rho_{\text{GE}}(E)dE=N.
\end{equation}
We determine the parameter $R_{0}$\ by requiring that the mean level density
is $1/D$ so that%
\begin{equation}
\frac{1}{N}%
{\displaystyle\int_{-\infty}^{\infty}}
\left[  \rho_{\text{GE}}(E)\right]  ^{2}dE=\frac{1}{D}.
\end{equation}
This condition yields
\begin{equation}
R_{0}=\frac{16N}{3\pi^{2}}D.
\end{equation}
Substituting (17) into (8) we obtain the following expression for the level
density of the superstatistical ensemble%
\begin{equation}
\rho_{\text{SE}}(E,\alpha)=%
{\displaystyle\int_{0}^{3\pi^{2}\left\vert E\right\vert /(16N)}}
f(D,\alpha)\rho_{\text{GE}}(E,D)dD.
\end{equation}
We could not solve this integral analytically. We evaluated it numerically for
different values of $\alpha$. The results of calculation are shown in Fig. 1.
The figure shows that the level density is symmetric with respect to $E=0$ for
all values of $\alpha$ and has a pronounced peak at the origin. However, the
behavior of the level density for finite $\alpha$ is quite distinct from the
semicircular law. It has a long tail whose shape and decay rate both depend on
the choice the parameter distribution $f(D)$. This behavior is similar to that
of the level density of mixed system modelled by a deformed random-matrix
ensemble \cite{bertuola1}.

\section{NEAREST-NEIGHBOR-SPACING DISTRIBUTION}

The NNS distribution is probably the most popular characteristic used in the
analysis of level statistics. In principle, it can be calculated once the
joint-eigenvalue distribution is known. The superstatistics generalization of
NNS distribution for an ensemble belonging to a given symmetry class is
obtained by substituting the NNS distribution of the corresponding Gaussian
ensemble $P_{\text{GE}}(s,D)$ for $\sigma^{(G)}(E,D)$\ in (7) and integrating
over the local mean level spacing $D$
\begin{equation}
P_{\text{SE}}(s)=\int_{0}^{\infty}f(D)P_{\text{GE}}(s,D)dD.
\end{equation}
Till now, no analytical expression for the NNS distribution could be derived
from RMT. What we know is that this distribution is very well approximated by
the Wigner surmise \cite{mehta}. We shall obtain superstatistics for NNS
distribution for systems with orthogonal and unitary symmetries by assuming
that the corresponding Gaussian ensembles have Wigner distributions for the
nearest-neighbor spacings.

Equation (22) yields the following relation between the second moment
$\left\langle D^{2}\right\rangle $ of the local-spacing distribution $f(D)$
and the second moment $\left\langle s^{2}\right\rangle $ of the spacing
distribution $P_{\text{SE}}(s)$:
\begin{equation}
\left\langle D^{2}\right\rangle =\frac{\left\langle s^{2}\right\rangle
}{\left\langle s^{2}\right\rangle _{\text{GE}}},
\end{equation}
where $\left\langle s^{2}\right\rangle _{\text{GE}}$ is the mean square
spacing for the corresponding Gaussian ensemble. For the distribution in Eq.
(13), one obtains
\begin{equation}
\left\langle D^{2}\right\rangle =\frac{G_{03}^{30}\left(  \left.  \alpha
^{3}\right\vert 0,\frac{1}{2},1\right)  G_{03}^{30}\left(  \left.  \alpha
^{3}\right\vert 0,\frac{3}{2},2\right)  }{\left[  G_{03}^{30}\left(  \left.
\alpha^{3}\right\vert 0,1,\frac{3}{2}\right)  \right]  ^{2}}.
\end{equation}
Using the asymptotic behavior of the G-function, we find that $\left\langle
D^{2}\right\rangle \rightarrow1$ as $\alpha\rightarrow\infty$, while
$\left\langle D^{2}\right\rangle =2$ (as for the Poisson distribution) when
$\alpha=0$. For practical purposes, the expression in Eq.(24) can be
approximated with a sufficient accuracy by $\left\langle D^{2}\right\rangle
\approx1+1/(1+4.121\alpha)$. Thus, given an experimental or
numerical-experimental NNS distibution, one can evaluate the quantity
$\left\langle s^{2}\right\rangle $\ and estimate the corresponding value of
the parameter $\alpha$ by means of the following approximate relation
\begin{equation}
\alpha\approx0.243\frac{\left\langle s^{2}\right\rangle }{\left\langle
s^{2}\right\rangle -\left\langle s^{2}\right\rangle _{\text{GE}}}.
\end{equation}

\subsection{Orthogonal ensembles}

Systems with spin-rotation and time-reversal invariance belong to the
orthogonal symmetry class of RMT. Chaotic systems of this class are modeled by
GOE for which NNS is well approximated by the Wigner surmise
\begin{equation}
P_{\text{GOE}}(s,D)=\frac{\pi}{2D^{2}}s\exp\left(  -\frac{\pi}{4D^{2}}%
s^{2}\right)  .
\end{equation}
We now apply superstatistics to derive the corresponding NNS distribution
assuming that the local mean spacing distribution $f(D)$\ is given by Eq.
(13). Substituting (26) into (22), we obtain
\begin{equation}
P_{\text{SOE}}(s,\alpha)=\frac{\pi\alpha^{2}}{2D_{0}^{2}G_{03}^{30}\left(
\left.  \alpha^{3}\right\vert 0,\frac{1}{2},1\right)  }sG_{03}^{30}\left(
\left.  \alpha^{3}+\frac{\pi\alpha^{2}}{4D_{0}^{2}}s^{2}\right\vert -\frac
{1}{2},0,0\right)  ,
\end{equation}
where $D_{0}$\ is given by (14), while the suffix SOE stand for
Superstatistical Orthogonal Ensemble.

Because of the difficulties of calculating $G_{0,3}^{3,0}\left(  z\left\vert
b_{1},b_{2},b_{3}\right.  \right)  $ at large values of $z,$ we use (say for
$z>100$) the large $z$ asymptotic formula given in the Appendix to obtain
\begin{equation}
P_{\text{SOE}}(s,\alpha)\approx\frac{\pi}{2}s\frac{\exp\left[  -3\alpha\left(
\sqrt[3]{1+\frac{\pi s^{2}}{4\alpha}}-1\right)  \right]  }{\sqrt{1+\frac{\pi
s^{2}}{4\alpha}}},
\end{equation}
which clearly tends to the Wigner surmise for the GOE as $\alpha$ approaches
infinity. This formula provides a reasonable approximation for $P_{\text{SOE}%
}(s,\alpha)$\ at sufficiently large values of $s$ for all values of
$\alpha\neq0.$ In this respect, the asymptotic behavior of the
superstatistical NNS distribution is given by
\begin{equation}
P_{\text{SOE}}(s,\alpha)\backsim C_{1}\exp\left(  -C_{2}s^{2/3}\right)  ,
\end{equation}
where $C_{1,2}$ are constants, unlike that of the NNS distribution obtained by
Tsallis' non-extensive statistics \cite{toscano}, which asymptotically decays
according to a power law.

Figure 2 shows the evolution of $P_{\text{SOE}}(s,\alpha)$ from a Wigner form
towards a Poissonian shape as $\alpha$ decreases from $\infty$ to 0. This
distribution behaves similarly but not quite exactly as any member of the
large family of distributions. One of these is Brody's distribution
\cite{brody}, which is given by
\begin{equation}
P_{\text{Brody}}(s,\gamma)=a_{\gamma}s^{\gamma}\exp\left(  -a_{\gamma
}s^{\gamma+1}/(\gamma+1)\right)  ,a_{\gamma}=\frac{1}{\gamma+1}\Gamma
^{\gamma+1}\left(  \frac{1}{\gamma+1}\right)  .
\end{equation}
This distribution is very popur but essentially lacks a theoretical
foundation. It has been frequently used in the analysis of experiments and
numerical experiments. The evolution of the Brody distribution during the
stochastic transition is shown also in Fig.2. The Brody distribution coincides
with the Wigner distribution if $\gamma=1$\ and with Poisson's if $\gamma=0$.
On the other hand, the superstatistical distribution at $\alpha=0$ is slightly
different, especially near the origin. For example, one can use the
small-argument expression of Mejer's G-function to show that $\lim
_{\alpha\longrightarrow0,s\longrightarrow0}P_{\text{SOE}}(s,\alpha)=\pi/2$. In
the mid-way of the stochastic transition, the agreement between the two
distributions that is only qualitative. At small $s$, the superstatistical
distribution increases linearly with $s$ while the increase of the Brody
distribution is faster. The large $s$ behavior is different as follows from
Eqs. (29) and (30). The difference between the two distributions decreases as
they approach the terminal point in the transition to chaos where they both
coincide with the Wigner distribution.

The superstatistical NNS distribution for systems in the midway of a
stochastic transition weakly depends on the choice of the parameter
distribution. To show this, we consider other two spacing distributions, which
have previously been obtained using other superstatistics
\cite{supst,abul1,abul2}. The first is \ derived from the uniform
distribution, considered in the original paper of Beck and Cohen \cite{BC}.
The second is obtained for a $\chi^{2}$-distribution of the parameter $\eta$,
which is known to produce Tsallis' non-extensive theory. In the latter case,
we qualify the NNS distribution by the parameter $m=\frac{2}{q-1}-d-2$, where
$q$ is Tsallis' entropic index and $d$ is the dimension of the Hamiltonian
random matrix. This behavior is quite different from the conventional NNS
which are frequently used in the analysis of experiments and nuclear
experiments, namely Brody's and Izrailev's \cite{izrailev}. The latter
distribution is given by%
\begin{equation}
P_{\text{Izrailev}}(s,\lambda)=As^{\lambda}\exp\left(  -\frac{\pi^{2}\lambda
}{16}s^{2}-\frac{\pi}{4}\left(  B-\lambda\right)  s\right)  ,
\end{equation}
where $A$ and $B$\ are determined for the conditions of normalization and unit
mean spacing. Figure 3 demonstrates the difference between the
superstatistical and conventional distribution in the mid-way between the
ordered and chaotic limits. The figures compares these distributions with
parameters that produce equal second moments. The second moment of the Brody
distribution is given by
\begin{equation}
\left\langle s^{2}\right\rangle _{\text{Brody}}=\frac{\Gamma\left(  1+\frac
{2}{\gamma+1}\right)  }{\Gamma^{2}\left(  1+\frac{1}{\gamma+1}\right)  }.
\end{equation}
We take $\gamma=0.3$, calculate $\left\langle s^{2}\right\rangle
_{\text{Brody}}$ and use the corresponding expressions for the second moment
of the other distributions to find the value of their tuning parameters that
makes them equal to  $\left\langle s^{2}\right\rangle _{\text{Brody}}$. The
comparison in Fig. 3 clearly shows that, while the considered three
superstatistical distributions are quite similiar, they considerably differ
from  Brody's and Izrailev's distributions.

The superstatistical distribution $P_{\text{SOE}}(s,\alpha)$ can still be
useful at least when Brody's distribution does not fit the data
satisfactorily. As an example, we consider a numerical experiment by Gu et al.
\cite{gu} on a random binary network. Impurity bonds are employed to replace
the bonds in an otherwise homogeneous network. The authors of Ref. \cite{gu}
numerically calculated more than 700 resonances for each sample. For each
impurity concentration $p$, they considered 1000 samples with totally more
than 700 000 levels computed. Their results for four values of concentration
$p$ are compared with both the Brody and superstatistical distribution in
Figure 4. The high statistical significance of the data allows us to assume
the advantage of the superstatistical distribution for describing the results
of this experiment.

\subsection{Unitary ensembles}

Now we calculate the superstatistical NNS distribution for a mixed system
without time-reversal symmetry. Chaotic systems belonging this class are
modeled by GUE for which the Wigner surmise reads
\begin{equation}
P_{\text{GUE}}(s,D)=\frac{32}{\pi^{2}D^{3}}s\exp\left(  -\frac{4}{\pi D^{2}%
}s^{2}\right)  .
\end{equation}
We again assume that the local mean spacing distribution $f(D)$\ is given by
Eq. (13). The superstatistics generalization of this distribution is obtained
by substituting (33) into (22),
\begin{equation}
P_{\text{SUE}}(s,\alpha)=\frac{32\alpha^{3}}{\pi^{2}D_{0}^{3}G_{03}%
^{30}\left(  \left.  \alpha^{3}\right\vert 0,\frac{1}{2},1\right)  }%
s^{2}G_{03}^{30}\left(  \left.  \alpha^{3}+\frac{4\alpha^{2}}{\pi D_{0}^{2}%
}s^{2}\right\vert -1,-\frac{1}{2},0\right)  ,
\end{equation}
where $D_{0}$\ is given by (14). At large values of $z,$ we use the large $z$
asymptotic formula for the G-function to obtain
\begin{equation}
P_{\text{SUE}}(s,\alpha)\approx\frac{32}{\pi^{2}}s^{2}\frac{\exp\left[
-3\alpha\left(  \sqrt[3]{1+\frac{4s^{2}}{\pi\alpha}}-1\right)  \right]
}{\left(  1+\frac{4s^{2}}{\pi\alpha}\right)  ^{5/6}},
\end{equation}
which clearly tends to the Wigner surmise for the GUE as $\alpha$ approaches
infinity as in the case of a GOE.

Figure 5 shows the behavior of $P_{\text{SUE}}(s,\alpha)$ for different values
of $\alpha$ ranging from $0$ to $\infty$ (the GUE). As in the case of the
orthogonal universality, the superstatistical distribution is not exactly
Poissonian when $\alpha=0$. Using the small argument behavior of Mejer's
G-function, one obtains $\lim_{\alpha\longrightarrow0,s\longrightarrow
0}P_{\text{SOE}}(s,\alpha)=4/\pi$.

\section{TWO-LEVEL CORRELATION FUNCTION}

The two-level correlation function is especially important for the statistical
analysis of level spectra \cite{guhr}. It is also directly related to other
important statistical measures, such as the spectral rigidity $\Delta_{3}$ and
level-number variance $\Sigma^{2}$. These quantities characterize the
long-range spectral correlations which have little influence on NNS distribution.

The two-level correlation function $R_{2}(E_{1},E_{2})$ is obtained from the
eigenvalue joint distribution function $P_{\beta}^{(G)}(\eta,E_{1}%
,\cdots,E_{N})$ by integrating over all eigenvalues except two. It is usually
broken into a connected and disconnected parts. The disconnected part is a
product of two level densities. On the unfolded spectra, the corresponding
two-level correlation function can be written as \cite{mehta,guhr}
\begin{equation}
X_{2}\left(  \xi_{1},\xi_{2}\right)  =D^{2}R_{2}\left(  D\xi_{1},D\xi
_{2}\right)  .
\end{equation}
Here the disconnected part is simply unity and the connected one, known as the
two-level cluster function, depends on the energy difference $r=\xi_{1}%
-\xi_{2}$ because of the translation invariance. One thus writes
\begin{equation}
X_{2}(r)=1-Y_{2}(r).
\end{equation}
The absence of all correlation in the spectra in the case of the Poisson
regularity is formally expressed by setting all k-level cluster functions
equal 0, and therefore
\begin{equation}
X_{2}^{\text{Poisson}}(r)=1.
\end{equation}
We shall here consider the unitary class of symmetry. For a GUE, the two-level
cluster function is given by
\begin{equation}
Y_{2}^{\text{GUE}}(r)=\left(  \frac{\sin\pi r}{\pi r}\right)  ^{2}.
\end{equation}
The two-level correlation function for mixed system described by the
superstatistics formalism is given using Eqs. (7) and (26) by
\begin{equation}
X_{2}^{\text{SUE}}(r)=\frac{1}{\left\langle D^{-2}\right\rangle }\int
_{0}^{\infty}f(D)\frac{1}{D^{2}}X_{2}^{\text{GUE}}(\frac{r}{D})dD,
\end{equation}
where we divide by $\left\langle D^{-2}\right\rangle $ in order to get the
correct asymptotic behavior of $X_{2}(r)\rightarrow1$ as $r\rightarrow\infty$.
Unfortunately, we were not able to evaluate this integral analytically in a
closed form. The results of numerical calculation of $X_{2}^{\text{SUE}}(r)$
for $\alpha=0.5,1$ and $\infty$ (the GUE) are given in Fig 6.

\section{SUMMARY AND CONCLUSION}

We have constructed a superstatistical model that allows to describe systems
with mixed regular-chaotic dynamics within the framework of RMT. The
superstatistics arises out of a superposition of two statistics, namely one
described by the matrix-element distribution $\exp\left[  -\eta\text{Tr}%
\left(  H^{\dagger}H\right)  \right]  $\ and another one by the probability
distribution of the characteristic parameter $\eta$. The latter defines the
energy scale; it is proportional to the inverse square of the local mean
spacing $D$ of the eigenvalues. The proposed approach is different from the
usual description of mixed systems, which model the dynamics by ensembles of
deformed or banded random matrices. These approaches depend on the basis in
which the matrix elements are evaluated. The superstatistical expressions
depend on Tr$\left(  H^{\dagger}H\right)  $ which is invariant under base
transformation. The model represents the spectrum of a mixed system as
consisting of an ensemble of sub-spectra to which are associated different
values of the mean level spacing $D$. The departure of chaos is thus expressed
by introducing correlations between the matrix elements of RMT. Spectral
characteristics of mixed systems are obtained by integrating the respective
quantities corresponding to chaotic systems over all values of $D$. In this
way, one is able to obtain entirely new expressions for the NNS distributions
and the two-level correlation functions for mixed systems. These expressions
reduce to those of RMT in the absence of fluctuation of the parameter $D$,
when the parameter distribution is reduced to a delta function. They can be
used to reproduce experimental results for systems undergoing a transition
from the statistics described by RMT towards the Poissonian level statistics,
especially when conventional models fail. This has been illustrated by an
analysis of a high-quality numerical experiment on the statistics of resonance
spectra of disordered binary networks.

\section{APPENDIX}

For sake of completeness, we give in this appendix the definition of the
Meijer G-function as well as some of its properties, which have been used in
the present paper. Meijer's G-function is defined by
\begin{equation}
G_{p,q}^{m,n}\left(  z\left\vert
\begin{array}
[c]{c}%
a_{1},\cdots,a_{p}\\
b_{1},\cdots,b_{q}%
\end{array}
\right.  \right)  =\frac{1}{2\pi i}\int_{L}\frac{\prod_{j=1}^{m}\Gamma\left(
b_{j}+s\right)  \prod_{j=1}^{n}\Gamma\left(  1-a_{j}-s\right)  }{\prod
_{j=m+1}^{q}\Gamma\left(  1-b_{j}-s\right)  \prod_{j=n+1}^{p}\Gamma\left(
a_{j}+s\right)  }z^{-s}ds,
\end{equation}
where $0\leq n\leq p$ and $0\leq m\leq q$ while an empty product is
interpreted as unity. The contour $L$ is a loop\ beginning and ending at
$-\infty$ and encircling all the poles of $\Gamma\left(  b_{j}+s\right)
,j=1,\cdots,m$ once in the positive direction but none of the poles of
$\Gamma\left(  1-a_{j}-s\right)  ,j=1,\cdots,n$. Various types of contours,
existence conditions and properties of the G-function are given in
\cite{mathai}. The way by which integrals of the type considered in this paper
are expressed in terms of the G-functions are described in \cite{mathai1}.

The asymptotic behavior of Meijer's G-function, as $\left\vert z\right\vert
\rightarrow\infty$, is given by \cite{luke}
\begin{equation}
G_{p,q}^{m,n}\left(  z\left\vert
\begin{array}
[c]{c}%
a_{1},\cdots,a_{p}\\
b_{1},\cdots,b_{q}%
\end{array}
\right.  \right)  \sim\frac{(2\pi)^{\left(  \sigma-1\right)  /2}}{\sigma
^{1/2}}z^{\theta}\exp\left(  -\sigma z^{1/\sigma}\right)  ,
\end{equation}
where $\sigma=q-p>0$, and $\sigma\theta=\frac{1}{2}(1-\sigma)+\sum_{j=1}%
^{q}b_{j}-\sum_{j=1}^{p}a_{j}$. In particular, the G-function that appears in
this paper
\begin{equation}
G_{0,3}^{3,0}\left(  z\left\vert b_{1},b_{2},b_{3}\right.  \right)  =\frac
{1}{2\pi i}\int_{L}\frac{1}{\Gamma\left(  1-b_{1}-s\right)  \Gamma\left(
1-b_{2}-s\right)  \Gamma\left(  1-b_{3}-s\right)  }z^{-s}ds,
\end{equation}
has the following asymptotic behavior
\begin{equation}
G_{0,3}^{3,0}\left(  z\left\vert b_{1},b_{2},b_{3}\right.  \right)  \sim
\frac{2\pi}{\sqrt{3}}z^{(b_{1}+b_{2}+b_{3}-1)/3}\exp\left(  -3z^{1/3}\right)
.
\end{equation}

On the other hand, the small $z$ behavior of Meijer's G-function
\cite{wolfram} is given by%
\begin{multline}
G_{p,q}^{m,n}\left(  z\left\vert
\begin{array}
[c]{c}%
a_{1},\cdots a_{n},a_{n+1},\cdots,a_{p}\\
b_{1},\cdots b_{m},b_{m+1},\cdots,b_{q}%
\end{array}
\right.  \right)  =%
{\displaystyle\sum\limits_{k=1}^{m}}
\frac{%
{\displaystyle\prod\limits_{\substack{j=1\\j\neq k}}^{m}}
\Gamma\left(  b_{j}-b_{k}\right)
{\displaystyle\prod\limits_{j=1}^{n}}
\Gamma\left(  1-a_{j}-b_{k}\right)  }{%
{\displaystyle\prod\limits_{j=n+1}^{p}}
\Gamma\left(  a_{j}-b_{k}\right)
{\displaystyle\prod\limits_{j=m+1}^{q}}
\Gamma\left(  1-b_{j}-b_{k}\right)  }\\
z^{b_{k}}\left[  1+\frac{%
{\displaystyle\prod\limits_{j=1}^{p}}
\left(  1-a_{j}-b_{k}\right)  }{%
{\displaystyle\prod\limits_{j=1}^{n}}
\left(  1-b_{j}-b_{k}\right)  }\left(  -1\right)  ^{-m-n+p}z+\cdots\right]  .
\end{multline}
Thus, the leading term in the expansion of $G_{0,3}^{3,0}\left(  z\left\vert
b_{1},b_{2},b_{3}\right.  \right)  $ in powers of $z$ is given by%
\begin{equation}
G_{0,3}^{3,0}\left(  z\left\vert b_{1},b_{2},b_{3}\right.  \right)
\approx\Gamma\left(  b_{2}-b_{1}\right)  \Gamma\left(  b_{3}-b_{1}\right)
z^{b_{1}},
\end{equation}
where $b_{1}$ is the smallest of $b_{i}$.

The implementation of Meijer's G-function in Mathematica \cite{wolfram}
constitutes an additional utility for analytic manipulations and numerical
computations involving this special function.

{\Large \pagebreak}

\bigskip Figure Caption (Color on Line)

Fig. 1. Level density for \ superstatistical orthogonal ensembles with
parameter $\alpha=0.2,1$ and $\infty$\ (the GOE limit).

Fig. 2. Evolution of NNS distributions obtained by the superstatistics method
for systems undergoing a transition from the GOE statistics to the Poissonian,
compared with the Brody's distributions.

Fog.3. Comparison between the superstatistical and conventional NNS
distributions having equal second moments.

Fig. 4. NNS distributions of geometrical resonances in random network,
calculated by Gu et al. \cite{gu} compared with the Brody and superstatistical distributions.

Fig. 5. NNS distributions obtained by the superstatistics method for systems
undergoing a transition from the GUE statistics to the Poissonian

Fig. 6. Two-level correlation functions obtained by the superstatistics method
for systems undergoing a transition from the GUE statistics to the Poissonian.

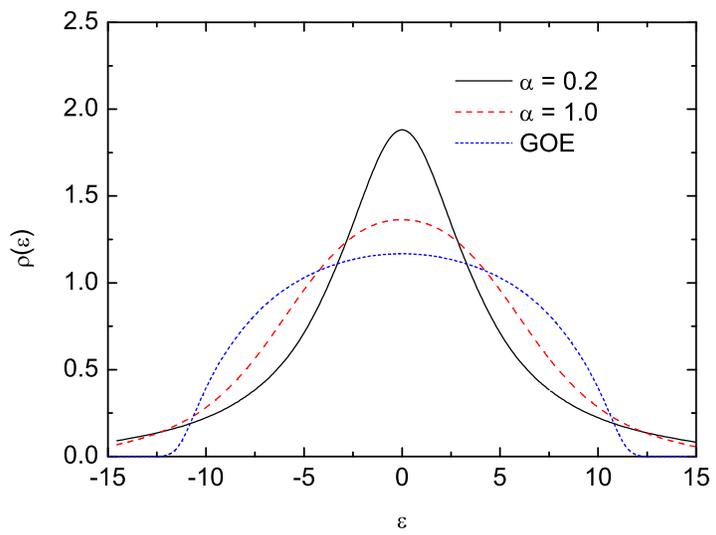
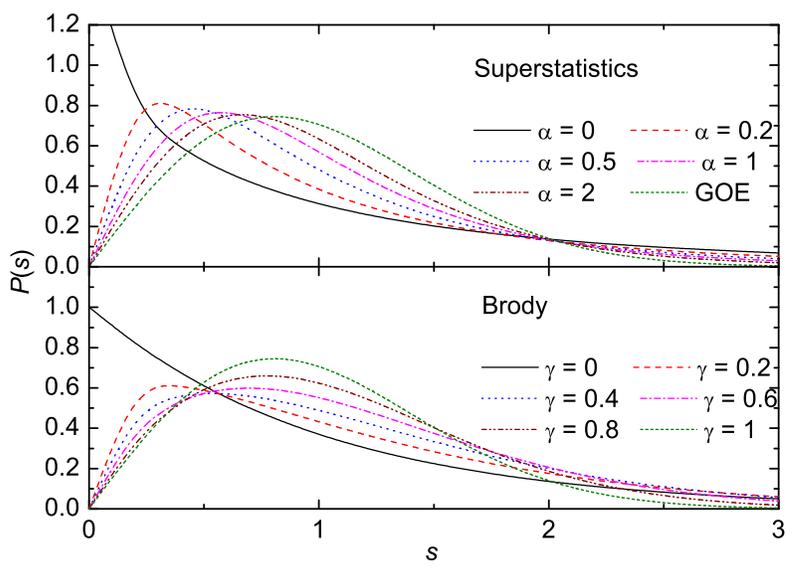
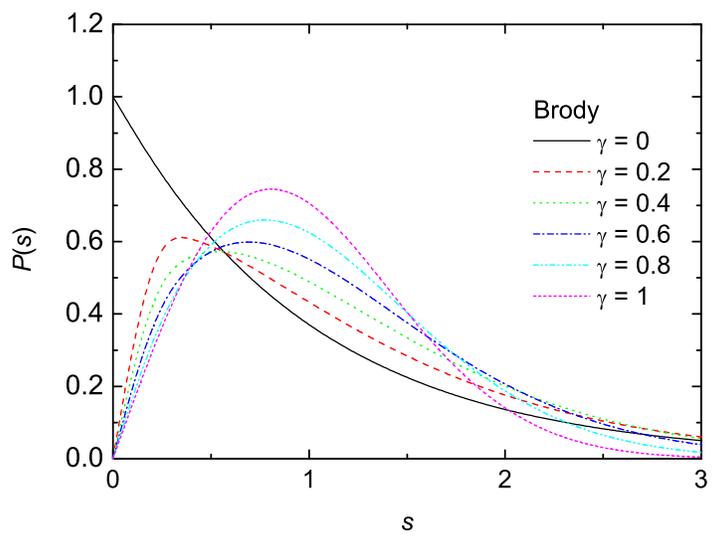
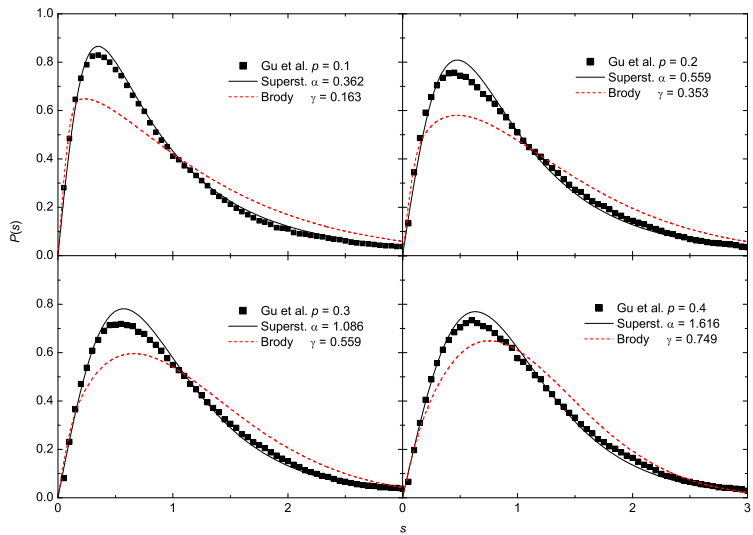
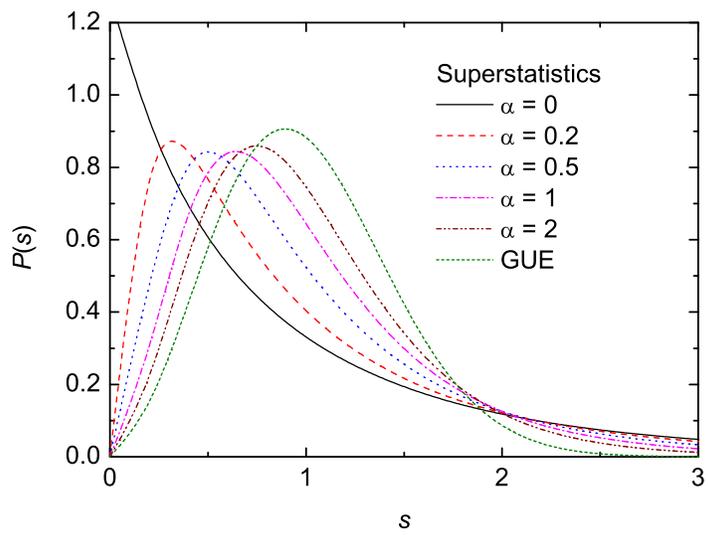
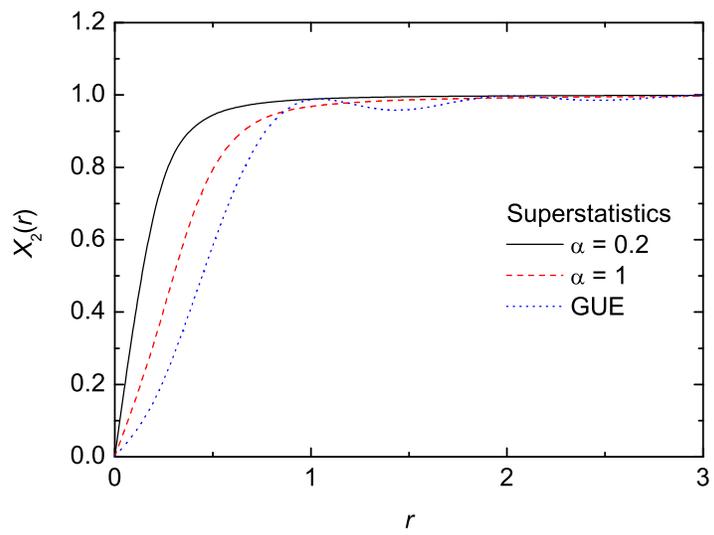
\end{document}